OPEN

# Proteomic and metagenomic insights into prehistoric Spanish Levantine Rock Art

Clodoaldo Roldán[1], Sonia Murcia-Mascarós[1], Esther López-Montalvo[2], Cristina Vilanova[3] & Manuel Porcar[3,4]

**The Iberian Mediterranean Basin is home to one of the largest groups of prehistoric rock art sites in Europe. Despite the cultural relevance of prehistoric Spanish Levantine rock art, pigment composition remains partially unknown, and the nature of the binders used for painting has yet to be disclosed. In this work, we present the first omic analysis applied to one of the flagship Levantine rock art sites: the Valltorta ravine (Castellón, Spain). We used high-throughput sequencing to provide the first description of the bacterial communities colonizing the rock art patina, which proved to be dominated by Firmicutes species and might have a protective effect on the paintings. Proteomic analysis was also performed on rock art microsamples in order to determine the organic binders present in Levantine prehistoric rock art pigments. This information could shed light on the controversial dating of this UNESCO Cultural Heritage, and contribute to defining the chrono-cultural framework of the societies responsible for these paintings.**

Spanish Levantine rock art is a unique pictorial expression in Prehistoric Europe due to the naturalism of depictions, the narrative component of scenes, and the vast distribution of rock-art shelters in the Iberian Mediterranean basin (Fig. 1a,b). Despite the historical and aesthetic value of Levantine rock art and its unprecedented distribution, its chrono-cultural framework is still unclear, due to the difficulties in radiocarbon dating of pigments[1]. Consequently, disparate chronological hypotheses have been proposed and, recently, new arguments have fuelled the debate, resulting in two alternative hypotheses on the chrono-cultural framework for Levantine rock art: one hypothesis proposes a Neolithic origin linked to the Neolithisation process and the spread of the "Neolithic package" in the Iberian Peninsula (from the 6$^{th}$ to the 3$^{rd}$ millennium cal. BC)[2,3] while the other argues a Mesolithic affiliation, based mainly on the thematic content of scenes[4,5], since big game hunting is one of the most frequent activities portrayed in the Levantine graphic tradition.

Besides dating, we also lack information on the sequence of technical gestures (*chaîne opératoire*) of Levantine pigments. Most archaeological research has focused on the characterization of raw materials, rather than the technical processes involved in executing the paintings. This approach is justified by the assumption that Levantine pigments were simple solutions/suspensions rather than complex mixtures, since components such as proteins or lipids, which could act as binders in mixtures, had not been detected at that point. Recently, we developed a new protocol including archaeobotanical microanalysis and experimental archaeology that enabled us to detect binders used to prepare Levantine charcoal pigments; however, we were unsuccessful in identifying them physico-chemically by *in situ* energy dispersive X-ray spectrometry (EDXRF) and Raman spectroscopy[6].

In order to improve the characterization of Levantine pigment composition we also need a better understanding of the microbial communities associated to the rock art patina to assess their putative role in the preservation of the raw materials and binders used. Microbial activity is central to rock-art preservation and degradation[7–9]. In particular, microorganisms are known to alter the composition of the mineral patina covering rock art paintings. For instance, the production of oxalic acid and other organic acids as a consequence of microbial metabolism is linked with the reinforcement of the patina, whereas microbial solubilisation of carbonates results in the progressive degradation of the patina. Despite the important implications of microorganisms on rock art conservation, the high-throughput analysis of microbial communities associated to prehistoric rock art has not been reported to date, despite the impressive development of metagenomic sequencing techniques.

[1]Materials Science Institute of the University of Valencia (ICMUV), Catedrático José Beltrán 2, 46980, Paterna, Valencia, Spain. [2]UMR 5608 TRACES, French National Center for the Scientific Research (CNRS), University of Toulouse 2-Jean Jaurès. 5, Allée Antonio Machado, 31058, Toulouse, France. [3]Darwin Bioprospecting Excellence, SL., Parc Cientific Universitat de València, 46980, Paterna, Valencia, Spain. [4]Institute for Integrative Systems Biology (I2SysBio, Universitat de València-CSIC). Parc Cientific Universitat de València, 46980, Paterna, Valencia, Spain. Correspondence and requests for materials should be addressed to S.M.-M. (email: sonia.mascaros@uv.es)

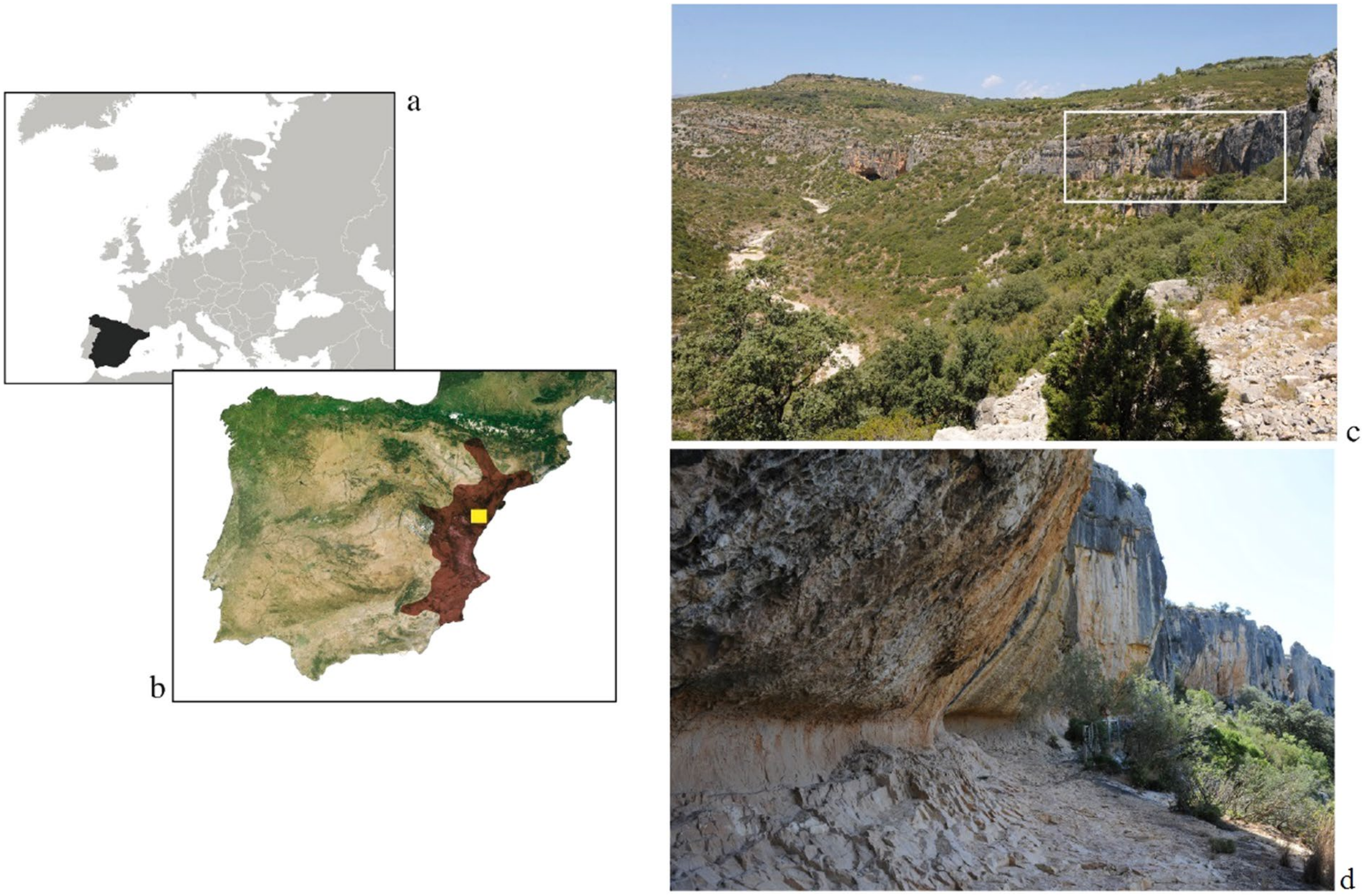

**Figure 1.** (**a**) Geographic location of the La Saltadora shelter (Coves de Vinromà, Castellón, Spain*). (**b**) The geographic zone of influence of Spanish Levantine rock art is shown in red**. (**c**) The location of La Saltadora in the Valltorta ravine is indicated by the white rectangle. (**d**) View of the shelters located in the Northern area of La Saltadora site. *The world map (**a**) (https://commons.wikimedia.org/wiki/File:BlankMap-World6.svg) is licensed under the Creative Commons Public Domain license. (https://creativecommons.org/publicdomain/mark/1.0/). The images are freely available for re-publication or re-use, including commercial purposes. **We acknowledge the use of data products or imagery from the Land, Atmosphere Near real-time Capability for EOS (LANCE) system operated by the NASA/GSFC/Earth Science Data and Information System (ESDIS) with funding provided by NASA/HQ.

In order to shed light on the composition and elaboration process of Levantine pigments, as well as on the variables involved in their stability and preservation, we present the first omic analysis of Levantine rock art. To do so, the case study reports on the red pigments of the Coves de la Saltadora rock-art shelter, located in the Valltorta ravine (Castellón, Spain) (Fig. 1c,d). This is one of the most outstanding Levantine rock art sites in the Iberian Mediterranean basin due to the number of motifs and the richness of styles and themes depicted. Red pigments from Saltadora shelters have previously been characterized *in situ* by means of portable EDXRF, which have revealed the presence of an iron-based pigment under a calcium-based superficial patina[10,11], in accordance with Levantine pictorial tradition[12–16].

The complete omic analysis reported here includes both a high-throughput sequencing of the bacterial communities colonizing the patina and a proteomic approach aiming to identify organic binders by means of analyzing the peptides present in the prehistoric pigments. Unveiling the composition of the microbial communities associated to the paintings has important implications for conservation strategies, while identification of peptidic profiles of Levantine pigments could be central to dating this prehistoric rock art. In this respect, binders could be used as chrono-cultural markers, which would provide insights into the framework of the individuals who authored the paintings.

## Results

Omic analyses were performed on three samples from Levantine motifs (CSI-01, CSV-01 and CSV-02), three samples from non-pigmented areas of the shelter close to the figures (CSI-04, CSV-03 and CSVI-01), three samples from non-pigmented areas of the shelter far from the figures (CSI-02, CSI-03 and CSVII-01) and two samples from surrounding rocks (COL and CRU) (Fig. 2, Table 1). The omic analysis of samples from Coves de la Saltadora was challenging due to the very small sample that was scraped off the wall of the shelter, especially in the pigmented areas, from which less than 15–20 mg was recovered. Despite this limitation, moderate amounts of metagenomic DNA were isolated from samples CSI-01, CSI-04, CSVI-01, COL, and CRU; and a significant number of proteins were also obtained from samples CSI-01, CSV-01, CSV-02, CSI-04, CSVI-01, CSI-02, CSI-03, and CSVII-01.

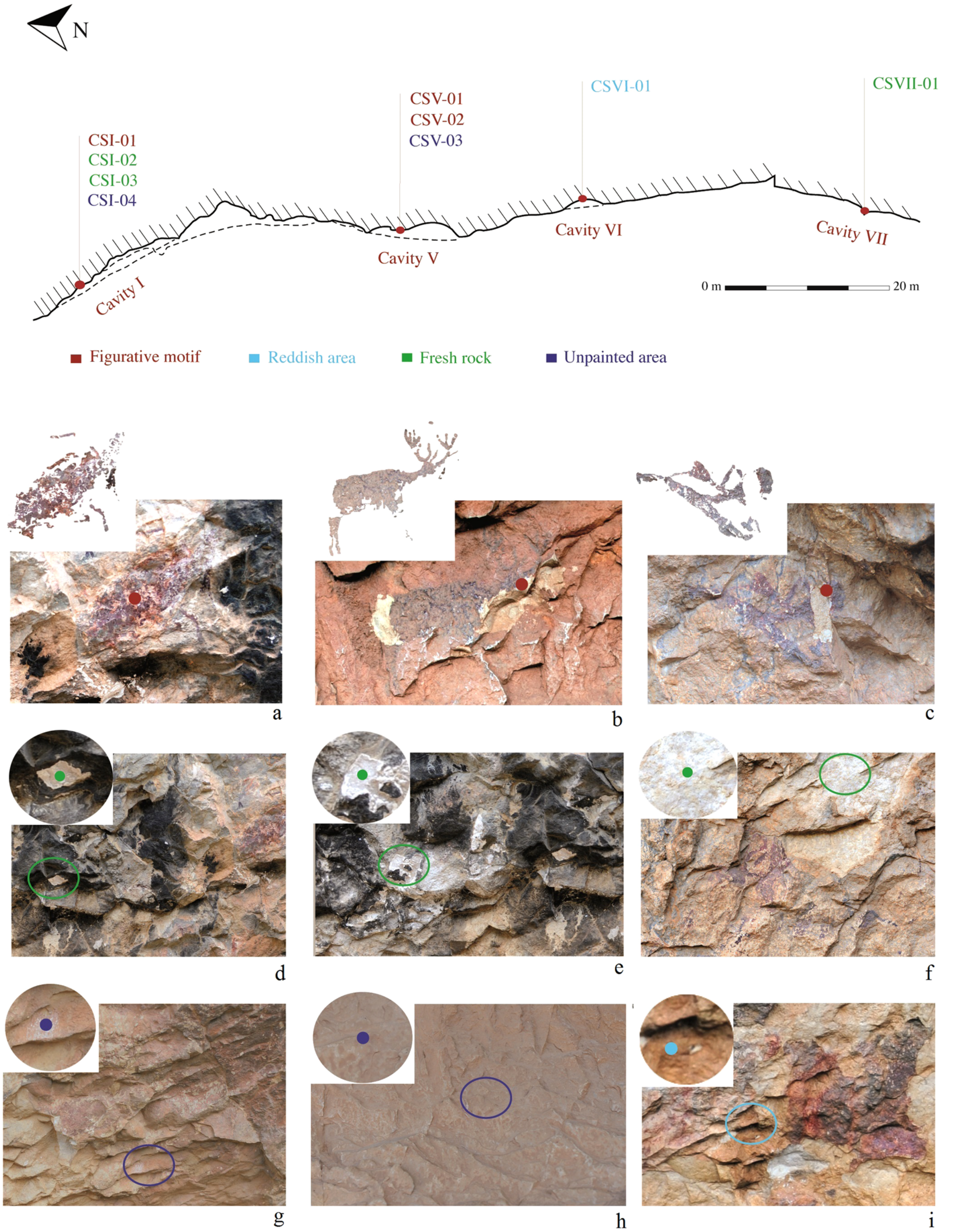

**Figure 2.** Floor plan of the Northern shelters of La Saltadora showing the distribution of the sampled figures and the control points. (**a**) Male goat (CSI-01). (**b**) Deer (CSV-01). (**c**) Archer (CSV-02). (**d**) Fresh rock (CSI-02). (**e**) Fresh rock (CSI-03). (**f**) Fresh rock (CSVII-01). (**g**) Unpainted area (CSI-04). (**h**) Unpainted area (CSV-03). (**i**) Reddish area (CSVI-01).

For the first time, massive 16S rRNA sequencing has unveiled the complete taxonomic diversity of bacteria colonizing a rock-art shelter. Some 200 species-level OTUs (Operational Taxonomic Units, which can be understood as microbial species) were detected on average per sample, with three OTUs accounting for 30% of the reads in two out of the three samples analyzed. These dominant OTUs belonged to Firmicutes species from genera *Staphylococcus*, *Enterococcus*, and *Thermicanus*. Other OTUs displaying a notable abundance in the patina were identified as *Acinetobacter* sp., *Rubellimicrobium* sp., *Pseudomonas* sp., and other species from the Comamonadaceae family (Fig. 3A). *Enterococcus* sp., *Thermicanus* sp., and *Bacillus thermoamylovorans* were especially rich in sample CSI-01 (corresponding to a pigmented area). Indeed, the taxonomic profile of this sample was slightly different to those of samples CSI-04 and CSVI-01 (non-pigmented areas), as shown in Fig. 3B, although differences were not statistically significant. Rock art communities were strikingly different to those associated with surrounding rocks (samples COL and CRU), which proved less diverse (around 65 species-level OTUs were detected) and clearly dominated by cyanobacteria (mainly from families Phormidiaceae and Nostocaceae; Fig. 3A).

| Sample Name | Location | Description |
|---|---|---|
| CSI-01 | Shelter I | Red figurative motif (goat) |
| CSV-01 | Shelter V | Red figurative motif (deer) |
| CSV-02 | Shelter V | Red figurative motif (archer) |
| CSI-04 | Shelter I | Unpainted area |
| CSV-03 | Shelter V | Unpainted area |
| CSVI-01 | Shelter VI | Reddish patina |
| CSI-02 | Shelter I | Fresh rock |
| CSI-03 | Shelter I | Fresh rock |
| CSVII-01 | Shelter VII | Fresh rock |
| CRU | surrounding rocks | Natural patina |
| COL | surrounding rocks | Natural patina |

**Table 1.** Samples description.

| OTU ID | Phylum | Order | Family | Genus |
|---|---|---|---|---|
| OTU1 | Firmicutes | Bacillales | Staphylococcaceae | *Staphylococcus* |
| OTU2 | Cyanobacteria | Oscillatoriales | Phormidiaceae | *Phormidium* |
| OTU3 | Cyanobacteria | Nostocales | Nostocaceae | — |
| OTU4 | Firmicutes | Lactobacillales | Enterococcaceae | *Enterococcus* |
| OTU5 | Acidobacteria | RB41 | — | — |
| OTU6 | Planctomycetes | WD2101 | — | — |
| OTU7 | Cyanobacteria | Stigonematales | Rivulariaceae | *Calothrix* |
| OTU8 | Proteobacteria | Pseudomonadales | Moraxellaceae | *Acinetobacter* |
| OTU9 | Cyanobacteria | Nostocales | Nostocaceae | — |
| OTU10 | Proteobacteria | Burkholderiales | Comamonadaceae | — |
| OTU11 | Firmicutes | Bacillales | Thermicanaceae | *Thermicanus* |
| OTU12 | OD1 | — | — | — |
| OTU13 | Acidobacteria | RB41 | Ellin6075 | — |
| OTU14 | Firmicutes | Bacillales | Bacillaceae | *Bacillus* |
| OTU15 | Proteobacteria | Rhodobacterales | Rhodobacteraceae | *Rubellimicrobium* |
| OTU16 | Chloroflexi | AKIW781 | — | — |
| OTU17 | TM7 | — | — | — |
| OTU18 | Cyanobacteria | Nostocales | Scytonemataceae | — |
| OTU19 | Proteobacteria | Pseudomonadales | Pseudomonadaceae | *Pseudomonas* |
| OTU20 | Bacteroidetes | Flavobacteriales | Flavobacteriaceae | *Flavobacterium* |

**Table 2.** Taxonomic identification of the 20 most abundant OTUs detected by 16S rRNA sequencing.

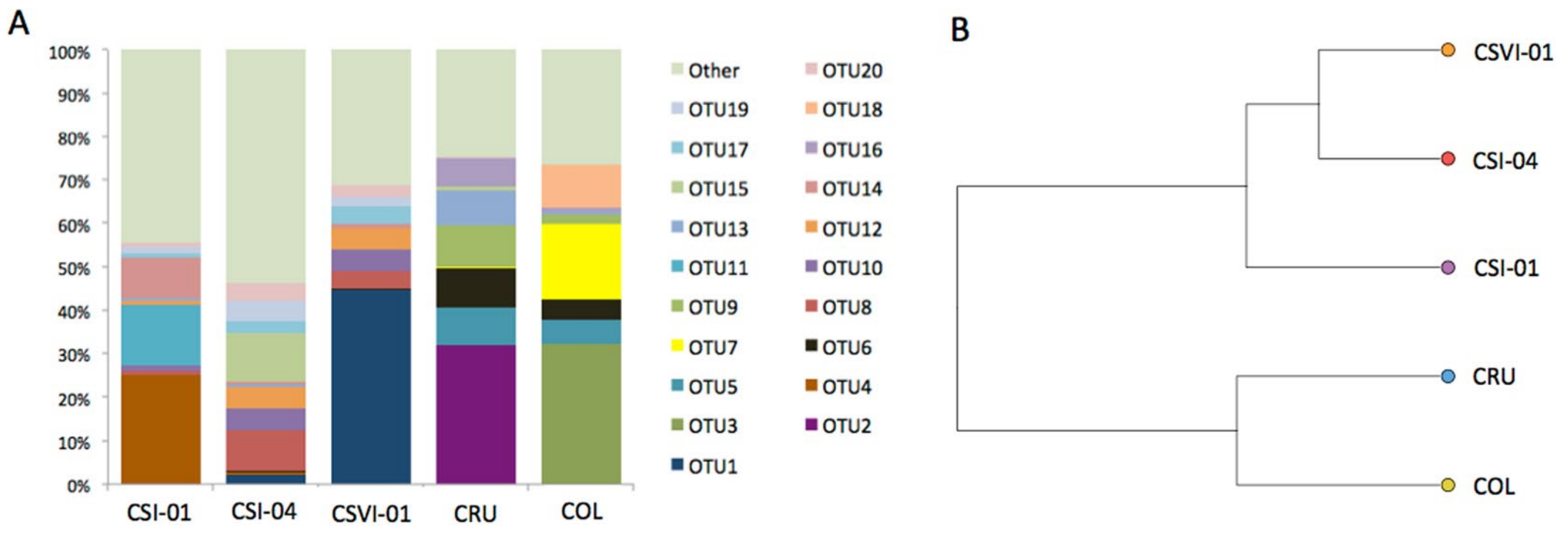

**Figure 3.** Taxonomic profiles of rock-art microbial communities obtained through 16S rRNA metagenomic sequencing. (**A**) Relative abundance of the 20 most abundant species-level OTUs detected in the analysis. OTUs are numbered in accordance with Table 2. (**B**) UPGMA clustering of the samples in accordance with their global taxonomic profile.

Proteomic analysis was able to detect around 100 different proteins in the pigmented areas of the patina (samples CSI-01, CSV-01, and CSV-02), and an average of 500 proteins in the other samples (non-pigmented areas in close proximity, and loose areas of the shelter). The vast majority of proteins detected matched with human keratins, which is a well-known signal of low protein content[17]. Apart from human keratins, a limited number of proteins were also identified by MS/MS in individual searches against the SWISSPROT database (Supplementary Data 1). For instance, bovine albumin and different human S100 proteins were found in all the samples coming from figurative motifs (CSI-01, CSV-01, and CSV-02); a range of bacterial ribosomal proteins were detected in sample CSV-02; and bovine casein and human S100 proteins, were detected in sample CSVI-01 (reddish patina). In order to increase the number of identified peptides per sample beyond those that had been sequenced and identified by MS/MS, we transferred peptide identifications to unsequenced peptides by matching their mass (m/z) and retention times (RT) (Supplementary Data 2). As a result, peaks corresponding to two peptides of bovine Alpha-S1-casein, with sequence FFVAPFPEVFGK and YLGYLEQLLR, were detected in more samples and thoroughly analyzed due to the implications that the presence animal casein has in the chrono-cultural debate surrounding Levantine paintings authorship. In order to compare the relative abundance of casein among all the samples, we calculated the proportion between the area of the peak corresponding to the casein peptide YLGYLEQLLR (quantified with a higher confidence) and the area of keratin peptides, the only peptides detected in all the samples. In order to avoid compositional effects due to the random variability of keratin peptides in terms of abundance, we used the sum of the area of 12 different peptides to minimize this effect on the relative quantification of casein. The Alpha-S1-casein peptide proved strikingly abundant in the pigmented samples CSI-01 and CSV-01 with respect to surrounding areas (Fig. 4A), with the only exception being sample CSVI-01 (reddish patina), which also proved rich in casein. A multiple alignment of the sequences of casein proteins corresponding to different species confirmed that the detected peptides were of bovine origin (Fig. 4B). Although the casein/keratins ratios should not be considered absolute, they can be used to estimate the relative distribution of the casein peptide in both pigmented and non-pigmented samples.

Additional characterization by Raman Spectroscopy has been performed on CSV-01 and CSV-02 microsamples, identifying the presence of hematite in the red pigment and a surface patina covering the motif itself, composed of oxalates, calcium carbonates or calcium sulphates (Supplementary Fig. S1).

## Discussion

Both the culture-independent microbial characterization and the proteomic strategies described in this work are unprecedented in studies of prehistoric rock art. Previous studies analyzed the microbial communities associated to Palaeolithic cave art in France[9] and northern Spain[18,19] by means of culture-dependent techniques and molecular approaches such as DGGE (denaturing gradient gel electrophoresis) analysis or 16S rRNA cloning and sequencing. The aforementioned studies reported a high abundance of Proteobacteria, Acidobacteria, and actinomycetes in the patinas. In contrast, the taxonomic profiles detected in our samples were dominated by Firmicutes, and proved more similar to the communities found in other locations subjected to conditions of lower humidity, such as Atlanterra and La Graja shelters in Southern Spain[20,21]. The latter locations, as in the samples analyzed here, were dominated by several species of genera *Bacillus*, *Paenibacillus*, and *Staphylococcus*. The prevalence of *Enterococcus* sp. and *Thermicanus* sp. detected in our samples has not been reported in rock art previously. The former is a highly adaptable generalist bacterial genus, present in both human- and non-human-associated environments[22], whereas *Thermicanus* sp is an uncommon microaerophilic thermophile first described from oxic soil in Egypt[23]. A diverse community of robust microorganisms is revealed by the combined presence of the abovementioned genera and other common environmental markers, such as *Pseudomonas* sp., or soil- and desert-associated genera such as *Rubellimicrobium* sp. The samples surrounding the shelter, subjected to conditions of higher humidity, were characterized by the abundance of cyanobacteria, similarly to the communities colonizing other types of cultural heritage, like stone monuments[24,25].

The implications of microbial communities in prehistoric rock art conservation are still controversial[8]. A range of fungi and bacteria have been linked to the biogenic production of oxalic acid, which generates a protective calcium oxalate patina on stone surfaces[26]. This is the case of bacteria from the genus *Bacillus*[7,27], found at high frequencies in our patina samples. However, some bacteria from the genus *Flavobacterium* (also detected in our samples, although at lower frequencies) are reported to deteriorate the oxalate patinas as a consequence of calcium carbonate solubilization[7]. A great deal of quantitative information is yielded by massive sequencing techniques, and it is tempting to propose a potential and novel role for microbial profiling, through deep sequencing, to identify frequencies and ratios of microbial oxalate producers and degraders that act as protectants or desestabilizers, respectively, of prehistoric pigments in cave art. For instance, in the well-preserved samples we analyzed, the *Bacilli:Flavobacteria* rate was high, suggesting that the bacterial communities in Coves de la Saltadora play a predominantly protective role. The sheer diversity of the bacteria unveiled by high-throughput sequencing of our samples is unprecedented, and further studies should establish the protective or deteriorating effect of most of the species detected. Notwithstanding, our results strongly suggest that microbiome determination can contribute to drawing up informed conservation strategies for Levantine rock art.

The detection of casein peptides of animal origin, showing perceptible differences in terms of relative abundance between the pigmented samples and the samples coming from other wall areas, could have dramatic implications from the historical point of view. Our finding is compatible with the use of animal milk as a binding agent in red pigments sampled from Coves de la Saltadora. Stockbreeding is only confirmed in the Iberian Peninsula from the 6th millennium cal. BC onwards, coinciding with the arrival and spread of the "Neolithic package"[28,29]. The main purposes of ovicaprine and bovine stockbreeding was believed to have been for human food-source and labour (animal draught), but several studies on lipid residues and slaughter profiles of domesticated ruminants in Northern Mediterranean have also recently proved the early exploitation of dairy products, probably representing a subsistence strategy for early farming communities[30]. In the vicinity of the Valltorta-Gassulla ravine, Neolithic sites provided domestic bovine (*Bos Taurus*) and ovicaprine (*Capra hircus* and *Ovis aries*) bones from the Early Neolithic levels (5400 cal. BC)[31,32]. Moreover, the location of some of the Neolithic sites in this area would indicate a function linked to the exploitation of mountain-pasture farming. The consumption of dairy products in the Iberian Mediterranean basin is attested through the identification of goat´s milk residues, by means of GC-IRMS, in several pottery vessels from Cova de l'Or (Beniarrès, Alicante, Spain)[33], dating from Early to Final Neolithic (5550 BC-2800 BC). The use of casein has also been evoked as a likely binder in several prehistoric and ethnographic rock art studies worldwide[34,35]. However, to date, only the GC analyses carried out on the Iberian Megalithic paintings of Dombate (Cabana de Bergantiños, A Coruña, Spain)[36] have suggested "cow butter" as a binder in pigment preparation.

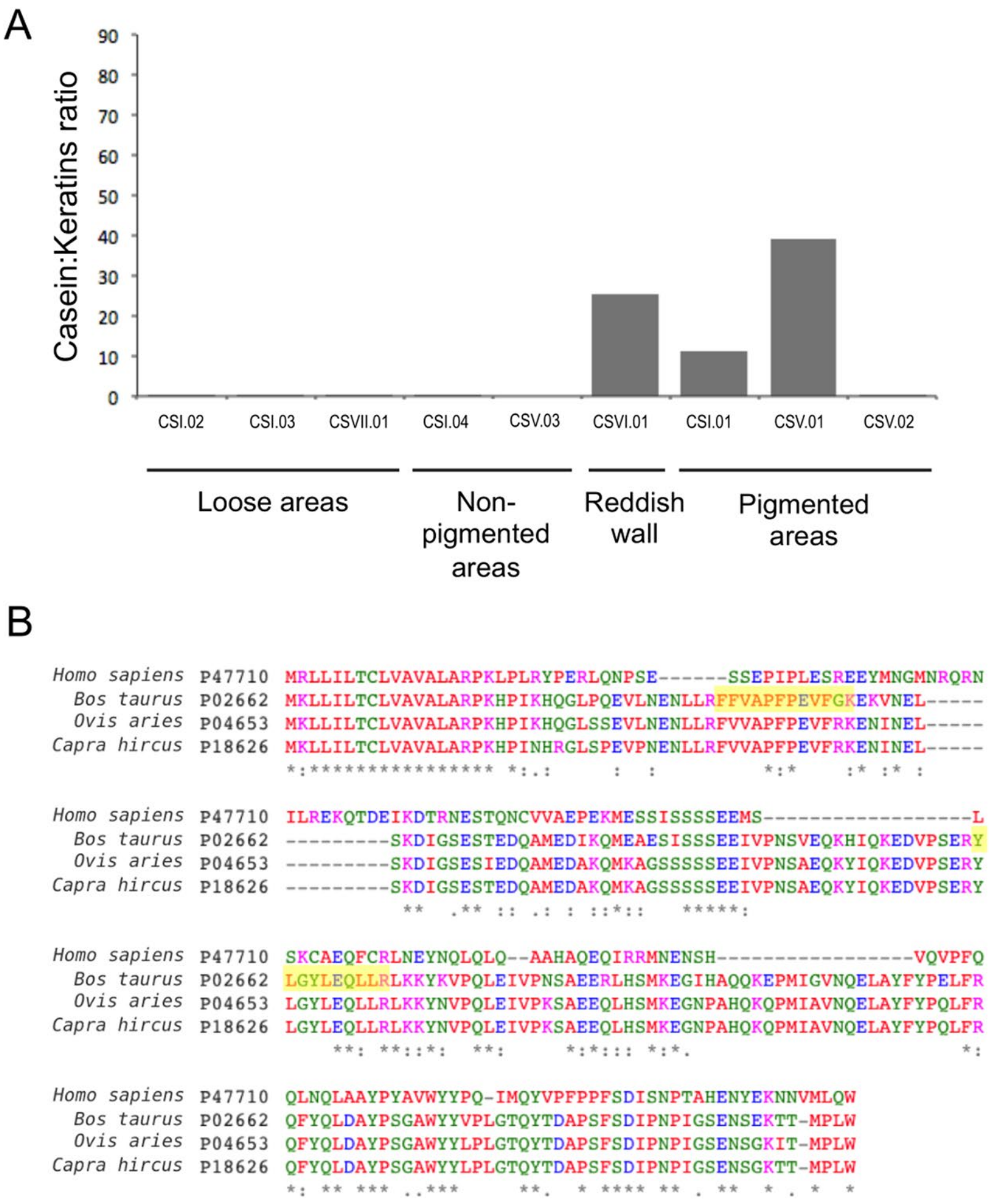

**Figure 4.** (**A**) Relative abundance of peptide YLGYLEQLLR in the samples analyzed with LC-MS/MS. Abundance was calculated with respect to a range of keratin peptides, as described in the Methods section. (**B**) Multiple alignments of Alpha-S1-casein proteins of different origins. The peptides detected in our analysis are highlighted in yellow.

From the cultural viewpoint, the use of milk is coherent with data provided by the regional archaeological record of the Valltorta-Gassulla ravine[37], while the experimental replica of pigments carried out in the laboratory showed that only the mixtures including animal fats provided the consistency and adherence required to paint on limestone[6]. Concerning the preparation process, we can hypothesize that powdered red ochre was mixed with liquid milk, with residual casein remaining after milk evaporation.

Besides the paintings themselves, we also detected the casein peptide YLGYLEQLLR in sample CSVI-01. This sample was taken from the Saltadora shelter number VI (Fig. 2), and the sampled surface presents a reddish patina that hides the precise localization of the red figures, and also has a very poor state of conservation. It is thus likely that the presence of casein in the sample is a consequence of either original or modern contamination from the originally defined figures that are no longer identifiable. On the other hand, the low casein/keratins ratio found in pigment sample CSV-02 could be attributed to the particularly small mass recovered during sampling from this red figurative motif.

A range of PTMs, including deamidation, were detected in the proteomic data. The analysis of PTMs was especially focused on glutamine deamidation rate, since it has been proposed as a marker of protein antiquity in the analysis of fossil materials[38,39]. However, we found no correlation between deamidation levels and putative, modern contaminants (i.e.: human keratins). This can be due to several factors, including the low number of peptides detected for each protein, and the influence of the preservational conditions on PTMs. Indeed, the use deamidation rate as a marker of antiquity is still controversial in this type of studies, since deamidation has been found to be highly dependent on several physicochemical factors[40].

There is still no direct evidence that discovered casein is not a ubiquitous contemporary contaminant. However, the detection of casein in a range of rock art samples worldwide[17,35] may support the hypothesis of casein as a common binder in prehistoric paintings, rather than as a universal contaminant.

Proteomic approaches have previously proved successful in order to characterize the organic binders present in art samples from a range of historical periods[17,41–44]. Albeit limited by sample size, our results show that the use of proteomics in rock art could also contribute to shedding light on technical and socioeconomic aspects of prehistoric societies. The presence of casein in Levantine pigments, and thus of milk, indicates a farming framework for Spanish Levantine rock art, in which a relatively abundant resource, cattle milk, would also have been used for technical or symbolic purposes. This observation is consistent with the excellent binding properties of animal fat and milk previously reported. Moreover, the presence of organic matter in Levantine pigments opens the prospect of obtaining radiocarbon dates, potentially enabling us to refine the chronological and cultural context of Levantine rock art.

Further, multidisciplinary efforts are needed to confirm whether the casein peptides we have identified in the present work are modern contaminants or, alternatively, molecular remains of the binders used by Iberian Neolithic societies.

## Methods

**Sampling.** Eleven sub-millimetric samples were collected from various locations throughout the La Saltadora rock-art shelter. Figure 2 shows the localization of the different sampled points. Sample material was scraped off the rock with a sterile scalpel by microbiologists under the supervision of the archeologists in charge. Samples were always handled under sterile conditions, using gloves and protease-treated materials in order to avoid contamination. Three samples (CSI-01, CSV-01 and CSV-02) came from different red figurative motifs; two samples (CSI-04 and CSV-03) came from unpainted zones near the red figures; one sample (CSVI-01) came from a reddish wall found within the shelter; and three samples (CSI-02, CSI-03 and CSVII-01) came from fresh rock surfaces of the shelter due to recent loosening of flake. These last six unpainted samples were collected for comparison with pigment samples. Finally, two samples (COL and CRU) were taken from natural patinas present in surrounding rocks outside the shelter (Table 1).

All the samples were handled under exactly the same conditions, in such a way that putative contaminants could not interfere in the detection of over-represented peptides in rock paintings (in comparison to control, unpigmented samples).

**DNA isolation and quantification.** DNA was extracted following the procedure previously described for rock art samples[45]. Briefly, cell lysis was achieved with lysozyme and proteinase K digestion, SDS treatment and freeze-thawing; and DNA was purified in silica gel membranes. Quantification was performed on a Qubit™ 3.0 Fluorometer following manufacturer's instructions.

**PCR amplification, library construction, and DNA sequencing.** A set of primers adapted to massive sequencing for the Illumina technology was used to amplify the 16S rRNA gene (V3-V4 hypervariable region[46]) from the metagenomic DNA in a PCR reaction. PCR reactions were performed with 30 ng of metagenomic DNA, 200 μM of each of the four deoxynucleoside triphosphates, 400 nM of each primer, 2.5 U of FastStart HiFi Polymerase, and the appropriate buffer with $MgCl_2$ supplied by the manufacturer (Roche, Mannheim, Germany), 4% of 20 g/mL BSA (Sigma, Dorset, United Kingdom), and 0.5 M Betaine (Sigma). Thermal cycling consisted of initial denaturation at 94 °C for 2 minutes followed by 35 cycles of denaturation at 94 °C for 20 seconds, annealing at 50 °C for 30 seconds, and extension at 72 °C for 5 minutes. Amplicons were combined in a single tube in equimolar concentrations. The pooled amplicon mixture was purified twice (AMPure XP kit, Agencourt, Takeley, United Kingdom) and the cleaned pool requantified using the PicoGreen assay (Quant-iT, PicoGreen DNA assay, Invitrogen). Subsequently, a sequencing library was prepared under a pair-end configuration following manufacturer's instructions, and sequencing on the Illumina MiSeq platform was performed at Life Sequencing SL facilities (Valencia, Spain). Sequencing statistics and rarefraction curves are shown in Supplementary Table S1 and Supplementary Figure S2, respectively.

**Protein isolation and digestion.** Rock art samples were resuspended in 15 μL Urea 8 M and 2 μL of ProteaseMax™ surfactant 0.2% (Promega) in 50 mM ABC (ammonium bicarbonate) in order to promote protein solubilization. The suspensions were vortexed for 2 min and agitated during 10 min at 140 rpm. Then, supernatants were transferred to a new tube, and mixed with 76.5 μL ABC 50 mM. Protein reduction was achieved by adding 1 μL DTT (dithiothreitol) 0.5 M and incubating the tubes at 56 °C for 20 min. Disulfide bonds were alkylated by adding 2.7 μL IAA (indoleacetic acid) 0.55 M and incubating the tubes in darkness during 15 min at room temperature. Finally, 1 μL of ProteaseMax™ surfactant 1% (Promega) in 50 mM ABC was added and protein digestion was carried by adding 250 ng of trypsin (Promega) and incubating the tubes at 37 °C for 3 h. The reaction was stopped by adding TFA (trifluoroacetic acid) to a final concentration of 0.5%.

Prior to LC-MS/MS, samples were dried, resuspended in 9 µL 2% ACN (acetonitrile) − 0.1% TFA, and centrifuged in order to separate peptides from degraded surfactant.

**Liquid chromatography and tandem mass spectrometry (LC–MS/MS).** Aliquots of 5 µl of each sample were loaded onto a trap column (NanoLC Column, 3 µ C18-CL, 350 µm × 0.5 mm Eksigent) and desalted with 0.1% TFA at 3 µl/min for 5 min. The peptides were then loaded onto an analytical column (LC Column, 3 µ C18-CL, 75 µ mx12 cm, Nikkyo) equilibrated in 5% acetonitrile 0.1% FA (formic acid). Elution was carried out with a linear gradient of 5 to 35% B in A for 30 min. (A: 0.1% FA; B: ACN, 0.1% FA) at a flow rate of 300 nl/min. Peptides were analyzed in a mass spectrometer nanoESIqQTOF (5600 TripleTOF, ABSCIEX).

Samples were ionized applying 2.8 kV to the spray emitter. Analysis was carried out in a data-dependent mode. Survey MS1 scans were acquired from 350–1250 m/z for 250 ms. The quadrupole resolution was set to 'UNIT' for MS2 experiments, which were acquired 100–1500 m/z for 50 ms in 'high sensitivity' mode. The following switch criteria were used: charge: 2+ to 5+; minimum intensity; 70 counts per second (cps). Up to 50 ions were selected for fragmentation after each survey scan. Dynamic exclusion was set to 15 s.

**Bioinformatic analysis of metagenomic data.** Metagenomic sequences were split taking into account the barcode introduced during the PCR reaction, providing two FASTQ files for each of the samples. Paired-end sequences were assembled with the PEAR package[47], then quality-filtered (Q20) using *fastx* tool kit version 0.013, primer-trimmed using *cutadapt* version 1.4.1, and length-trimmed (minimum 300 bp read length). The resulting FASTQ files were converted to FASTA files and UCHIME44 program version 7.0.1001 was used to remove chimeras arising during the amplification and sequencing step. After these processing steps, the taxonomic profiles were computed for each sample using software QIIME[48]. A pipeline consisting of the following steps was followed: first, sequences were clustered into OTUs with a 97% similarity threshold (corresponding to species-level OTUs) using an open-reference algorithm; second, a representative sequence was selected for each OTU; third, a taxonomic assignment was obtained for each representative sequence by means of BLAST searches against the GreenGenes database (versión 13_8); and finally, an OTU table was built by combining abundance and taxonomy data from each sample (Table 2). Data were represented and statistically analyzed with software MEGAN[49] and STAMP[50].

**Bioinformatic analysis of proteomic data.** Proteomic data were processed with software ProteinPilot (version 4.5.1, revision 2768; ParagonTM Algorithm 4.5.1.0, 2765; Sciex), which used MS/MS data for identifying peptide sequences and PTM (post-translational modification) sites against a protein sequence database. The default parameters were used to generate a peak list directly from 5600 TripleTof wiff files. The Paragon algorithm of ProteinPilot v 4.5 was used to search the SwissProt database (version 01-2016) with the following parameters: trypsin specificity, cys-alkylation, no taxonomy restriction, and the search effort set to "Thorough" mode. In order to increase the number of identified proteins, the pattern of each sequenced peptide was matched across different LC–MS/MS runs on the basis of mass and retention time. By using this approach, the software calculated the intensity for the identified peptides and also for features unidentified in their respective run, but with m/z and RT matching those of peptides identified in other runs[51]. The area under the chromatographic peak was calculated to assess peptide abundance.

Most of the chromatographic peaks corresponding to non-keratin proteins were notably weak, thus challenging their proper quantification and also the normalization of abundance data. To overcome this issue, peak identification and quantification were manually curated by supervising the chromatographs with PeakView software. For normalization purposes, the relative abundance of identified peptides was calculated as the proportion between the area of the chromatographic peak corresponding to each peptide and the sum of the areas of 12 peaks corresponding to keratin peptides that were detected in all the samples.

## Acknowledgements
This research was developed as a part of the project 'NEOSOCWESTMED' (no. 628428) under the Marie Curie Actions in the 7th Programme of the European Commission (FP7/2007–2013) and the "Projet Émergent 2016" of TRACES UMR 5608 of the French National Center for the Scientific Research (CNRS). The authors thank the General Directorate of Cultural Heritage of the Government of Valencia (Spain) for allowing us to carry out the –omic analysis of the pigments of Coves de la Saltadora. We also thank Josep Casabó and the cultural guides of the Museum of La Valltorta for assistance with our fieldwork and sharing their knowledge about Valltorta rock-art shelters. Finally, this work is dedicated to the memory of Mr. Eugenio Barreda, whose collaboration and support was decisive in the development of our research in Valltorta-Gassulla. S.T.T.L.

## Author Contributions
C.R., S.M.-M., E.L.-M. conceived the project; C.R., S.M.-M., E.L.-M. contributed resources; C.R., S.M.-M., E.L.-M., M.P., performed the sampling; S.M.-M., C.R., C.V., M.P. analyzed the data; and all authors contributed ideas, discussed the results, and wrote the manuscript.

## Additional Information
**Supplementary information** accompanies this paper at https://doi.org/10.1038/s41598-018-28121-6.

**Competing Interests:** The authors declare no competing interests.

**Publisher's note:** Springer Nature remains neutral with regard to jurisdictional claims in published maps and institutional affiliations.